\documentclass{emulateapj}
\usepackage{amssymb}
\usepackage{lscape}

\begin{document}
\newcommand{\lya}{Lyman~$\alpha$}
\newcommand{\lyb}{Lyman~$\beta$}
\newcommand{\degpoint}{\mbox{$^\circ\mskip-7.0mu.\,$}}
\newcommand{\minpoint}{\mbox{$'\mskip-4.7mu.\mskip0.8mu$}}
\newcommand{\secpoint}{\mbox{$''\mskip-7.6mu.\,$}}
\newcommand{\sqdeg}{\mbox{${\rm deg}^2$}}
\newcommand{\squig}{\sim\!\!}
\newcommand{\subsun}{\mbox{$_{\twelvesy\odot}$}}
\newcommand{\et}{{\it et al.}~}
\newcommand{\Rs}{{\cal R}}

\def\ltsima{$\; \buildrel < \over \sim \;$}
\def\simlt{\lower.5ex\hbox{\ltsima}}
\def\gtsima{$\; \buildrel > \over \sim \;$}
\def\simgt{\lower.5ex\hbox{\gtsima}}
\def\propsima{$\; \buildrel \propto \over \sim \;$}
\def\simprop{\lower.5ex\hbox{\propsima}}
\def\arcs{$''~$}
\def\arcm{$'~$}

\title{CONSTRAINTS FROM GALAXY-AGN CLUSTERING ON THE CORRELATION BETWEEN GALAXY AND BLACK HOLE MASS AT REDSHIFT $2\simlt Z\simlt 3$\altaffilmark{1}}

\author{\sc Kurt L. Adelberger\altaffilmark{2}}
\affil{Carnegie Observatories, 813 Santa Barbara St., Pasadena, CA, 91101}
                                                                                   
\author{\sc Charles C. Steidel}
\affil{Palomar Observatory, Caltech 105--24, Pasadena, CA 91125}
                                                                                   
\altaffiltext{1}{Based, in part, on data obtained at the W.M. Keck
Observatory, which is operated as a scientific partnership between
the California Institute of Technology, the University of California,
and NASA, and was made possible by the generous financial support
of the W.M. Keck Foundation.}
\altaffiltext{2}{Carnegie Fellow}

\begin{abstract}
We use the clustering
of galaxies around distant active-galactic nuclei (AGN) to derive an estimate
of the relationship between galaxy and black hole mass that obtained
during the ancient quasar epoch, at redshifts $2\simlt z\simlt 3$,
when giant black holes accreted much of their mass.
Neither the mean relationship nor its scatter differs
significantly from what is observed in the local universe,
at least over the ranges of apparent magnitude ($16\simlt G_{\rm AB}\simlt 26$)
and black-hole mass ($10^6\simlt M_{\rm BH}\simlt 10^{10.5}M_\odot$) 
that we are able to probe.
\end{abstract}

\submitted{Received 2005 April 24; Accepted 2005 May 25}
\shorttitle{CORRELATION OF GALAXY AND BLACK HOLE MASS AT $Z\sim 3$}
\shortauthors{Adelberger \& Steidel}

\section{INTRODUCTION AND RESULTS}
The study of black holes has been driven to the forefront of extragalactic
research by the recent discovery
of black holes as massive as
a billion suns inside nearby bulge galaxies.
Simple physical arguments (e.g., Silk \& Rees 1998)
suggest that these enormous objects should profoundly affect
the process of galaxy formation, a belief that is strengthened by 
the tight observed correlation between the masses of local galaxies and
their black holes~(Gebhardt et al. 2000; Ferrarese \& Merritt 2000).
Various theoretical models attempt to explain the existence of the correlation
with a wide range of physical processes.  Since these models make discordant
predictions for the evolution of the correlation over time, we decided
to test them by measuring the relationship between galaxy and black hole
mass in the distant past, at redshifts $2<z<3$.

\begin{figure}
\plotone{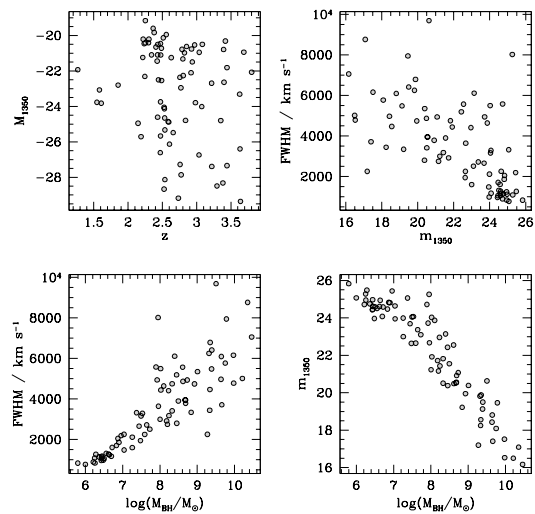}
\caption{\label{fig:check_mbh} 
Overview of the characteristics of the AGN in our sample.
{\it Upper left:} Redshifts and absolute AB magnitude
at rest-frame $1350$\AA.
The uncertainty in the AB magnitude
is $\simlt 0.2$ magnitudes for even our faintest objects (e.g.,
Steidel et al. 2003).
{\it Upper right:} Relationship between CIV line width and apparent AB magnitude
at rest-frame 1350\AA.
The uncertainty in line width
ranges from 10--20\%, and is dominated by
systematics (e.g., continuum placement) for the brightest AGN.
{\it Lower panels:} Relationship between CIV line width, $m_{1350}$,
and the resulting estimate of black-hole mass $M_{\rm BH}$.
The selection bias is severe in our AGN sample,
since (for example) we
deliberately targeted AGN that were bright and had broad emission lines.
These panels show the characteristics of our sample as selected, not
of a fair sample of high-redshift AGN.
}
\end{figure}

A novel approach (see, e.g., Kauffmann \& Haehnelt 2002)
let us use our existing surveys (Steidel et al. 2003;
Steidel et al. 2004)
of $\sim 1600$ galaxies at redshifts $1.5\simlt z\simlt 3$
to measure the dependence of galaxy mass $M_h$ on black hole mass $M_{\rm BH}$
over a 5-decade baseline of black hole mass, reaching masses roughly 1000 times
smaller than the limits of other surveys (e.g., Shields et al. 2003;
Walter et al. 2004; Croom et al. 2005) at similar redshifts.
After using the technique of Vestergaard (2002)
to estimate the masses of
the black holes that powered each of the 79 active-galactic nuclei (AGN) in our survey
(see figure~\ref{fig:check_mbh} and the appendix),
we estimated the typical halo mass for black holes in different
mass ranges by measuring how strongly the other
galaxies in our survey clustered around them.

\begin{figure}
\plotone{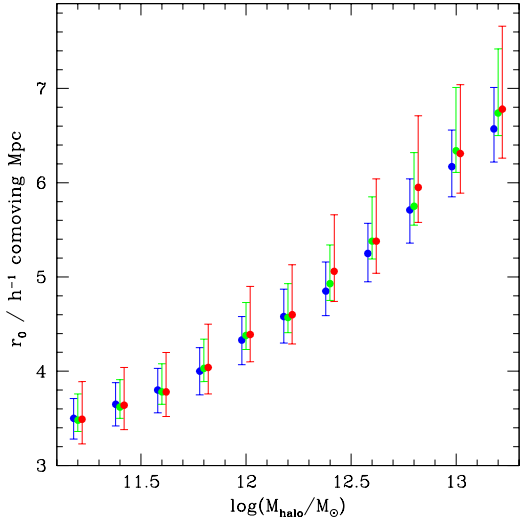}
\caption{\label{fig:r0_vs_mass} 
Relationship between cross-correlation length $r_0$
and implied mass of the black-holes' galaxies at redshifts $2\simlt z\simlt 3$.
Blue, green, and red points show the dependence of $r_0$
on $M_h$ in the GIF-LCDM simulation (Kauffmann et al. 1999)
at redshifts $z=2.97$, $2.74$, $2.32$.
Small offsets
have been added to the abscissae for clarity.  
The relationship is somewhat uncertain because
it depends on the assumed masses of the halos
that contain non-active galaxies.
Error bars show the $1\sigma$ uncertainty in the relationship that results 
from the uncertainty in the masses of the non-active galaxies.
}
\end{figure}

Adelberger \& Steidel (2005) describe our analysis in more detail.  
Briefly, we estimated
the cross-correlation length $r_0$
from the number of galaxy-AGN pairs with angular separation $60''<\theta<300''$
and comoving radial separation $\Delta Z<30h^{-1}$ Mpc
with the approach of Adelberger (2005), then used
the GIF-LCDM numerical simulation (Kauffmann et al. 1999) of structure formation in a
standard cosmological model to estimate from $r_0$ the total (dark matter plus baryon)
mean mass $M_h$ of the galaxies associated with black holes in each mass range.
The relationship between $r_0$ and $M_h$ depends on the redshift and on
the mass of the typical (non-active) galaxies in our survey, but 
the resulting systematic errors in $M_h$ are small compared
to the random errors (figure~\ref{fig:r0_vs_mass}).

\begin{figure}
\plotone{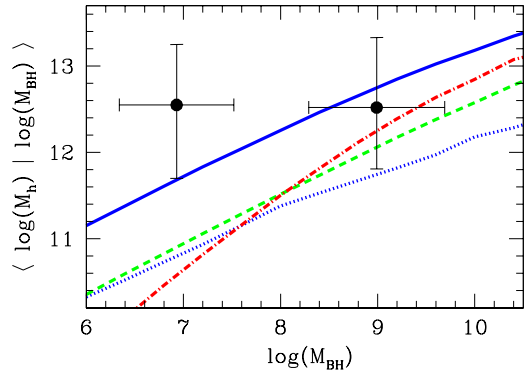}
\caption{\label{fig:emh_vs_mbh} 
Observed and expected relationship between black hole and halo mass
(in solar units) at redshift $z\sim 2.5$.  
The ordinate is $\langle\log(M_h)|\log(M_{\rm BH})\rangle$,
the mean value of $\log(M_h)$ for a given value of $\log(M_{\rm BH})$.
Points show our
observations.
Vertical error bars show the $1\sigma$ random uncertainty.
Horizontal error bars show the mean and rms value
of $\log(M_{\rm BH})$ for the two groups of black holes.  Lines show
theoretical predictions.
Solid blue line:  no evolution in
mean $M_{\rm BH}$--$M_h$ relationship,
negligible intrinsic scatter.
We adopt equation~6 of Ferrarese (2002) for the local
relationship, but her two alternatives fit our
observations comparably well.
Short-dashed blue line:  no evolution,
1 dex of scatter in $M_{\rm BH}$ at fixed $M_h$.
Green line: local relationship scaled by $(1+z)^{5/2}$.
Red line:  relationship at $z=3$ in
a supply-limited accretion model (Di Matteo et al. 2003).
Observations at large $M_{\rm BH}$ agree with any of these
scenarios, as has also been noted
by Shields et al. (2003) and Walter et al. (2004). Our small-$M_{\rm BH}$ data
help distinguish between them.
}
\end{figure}

We found galaxy-AGN cross-correlation lengths 
of $r_0=5.27^{+1.59}_{-1.36}$ for the $38$ AGN with $10^{5.8}<M_{BH}/M_\odot<10^8$ 
and $r_0=5.20^{+1.85}_{-1.16}$ for the $41$ with $10^8<M_{\rm BH}/M_\odot <10^{10.5}$.
The inferred relationship between $\log(M_{\rm BH})$ and $\langle\log(M_h)\rangle$
is shown in Figure~\ref{fig:emh_vs_mbh}.

If the predicted relationship between galaxy and black-hole mass
has the form $\log M_{\rm BH}  = f(\log M_h) +\epsilon$, with
$f$ a function to be specified and $\epsilon$ a random deviate, the expectation value of
$\log(M_h)$ for a given value of $\log(M_{\rm BH})$ 
follows from the elementary relationship
\begin{equation}
E(l_h|l_{\rm BH}) = \frac{\int_0^\infty dl_h l_h P(l_h) P(l_{\rm BH}|l_h)}{\int_0^\infty dl_h P(l_h) P(l_{\rm BH}|l_h)}
\label{eq:elhgivenlbh}
\end{equation}
where $l_h\equiv \log(M_h)$, $l_{\rm BH}\equiv \log(M_{\rm BH})$,
$P(l_h)$ is the distribution of $\log(M_h)$ measured
in the GIF-LCDM simulation and extrapolated with the appropriate
Press-Schechter (1974) formula,
and $P(l_{\rm BH}|l_h)$ is the distribution of $\log(M_{\rm BH})$ at
fixed galaxy mass, which depends on $f$ and on the characteristics of the random
variable $\epsilon$.  Solving equation~\ref{eq:elhgivenlbh} numerically
for different functions $f$
under the assumption that $\epsilon$ has a Normal distribution with rms $\sigma_\epsilon$,
we find the theoretical tracks shown in Figure~\ref{fig:emh_vs_mbh}.

The solid blue line is for a $M_{\rm BH}$--$M_h$ relationship
identical to the one observed locally,
$\log(M_{\rm BH}/10^7 M_\odot) = 1.65\log(M_h/10^{12}M_\odot) + \epsilon$
(Ferrarese 2002).
For this line we assumed $\sigma_\epsilon=0.5$, roughly the expected error
in our black hole masses (Vestergaard 2002).  The line therefore assumes negligible
intrinsic scatter in the correlation.  It fits the data well.

The other lines show that alternative relationships 
in the literature generally
provide a worse fit.
The green line results from scaling the ratio
of black hole to galaxy mass by $(1+z)^{5/2}$,
as advocated by many 
semi-analytic models (e.g.,
Haehnelt, Natarajan, \& Rees 1998; Wyithe \& Loeb 2002;
Volonteri, Haardt, \& Madau 2003).
The red line shows the redshift $z=3$ prediction
$M_{\rm BH}/M_\odot =6.2\times 10^7 (M_h/10^{12}M_\odot)^{1.033}$
of a model in which black holes accrete a fixed fraction
of the total gas mass in each merger (Di Matteo et al. 2003).
The dashed blue line assumes that the mean $M_{\rm BH}$--$M_h$ 
relationship is the same as observed locally but that its
intrinsic scatter has increased to $1.0$ dex.
Increasing the scatter decreases the typical mass 
of galaxies that contain black holes of a given mass.
This is because galaxies with low masses are much
more common than galaxies with high masses;
when the scatter in
the $M_{\rm BH}$--$M_h$ relationship is big, the largest black holes
are more likely to reside in low mass galaxies with unusual
ratios of $M_{\rm BH}/M_h$ than in high mass galaxies with normal ratios.
The clustering of galaxies around AGN would therefore be far weaker than we observe
if there were no relationship at all between $M_{\rm BH}$ and $M_h$.

A $\chi^2$ test suggests that the three alternatives to the
no-evolution model
($(1+z)^{5/2}$ scaling, supply limited accretion, large $\sigma_\epsilon$)
disagree with the observations at the 90--95\% level.  They can therefore be
considered marginally consistent with our present data, although
the odds are against them and more
extreme evolution from the local relationship (e.g., Haehnelt \& Rees 1993)
can be ruled out
with high significance.

The apparent lack of evolution in the
$M_{\rm BH}$--$M_h$ correlation seems consistent with models
in which the correlation results from active feedback from the black hole.
In these models the black hole mass is pinned near the maximum allowed
by its halo at all times.  If this maximum is set by the escape velocity
at a fixed proper radius from the black hole, it will
not depend strongly on redshift.
One might object that black holes are able to enter the quasar
phase in these models only because
their masses have temporarily fallen below the maximum
allowed by their growing halos, and so the most
luminous AGN should never lie on the correlation.
As long as the quasar phase occurs near
the end of the accretion, however (e.g., Hopkins et al. 2005), the black hole should have nearly
achieved its equilibrium mass.  In any case, a slight decrease
in $M_{\rm BH}$ at fixed $M_h$ would make the predictions fit
our data even better.

\bigskip
\bigskip
It is a pleasure to acknowledge several interesting conversations with
L. Ho, L. Hernquist, L. Ferrarese, J. Kollmeier, and S. White.
This work would not have been possible without the efforts
of our collaborators D. Erb, M. Pettini, N. Reddy, \& A. Shapley.
Bob Becker, Richard White, and
Michael Gregg generously shared their spectrum of FBQS0933+2845.
KLA was supported by
a fellowship from the Carnegie Institute of Washington; CCS
was supported by grant AST 03-07263
from the National Science Foundation and by a grant from the Packard foundation.
We are grateful that the people of Hawaii allow astronomers
to build and operate telescopes on the summit of Mauna Kea.

\appendix
\bigskip
\bigskip
\section{TECHNICAL DETAILS}
\bigskip
\medskip

The general population of galaxies tends to cluster more strongly
around individual galaxies with larger masses.  We exploit this effect
to estimate the masses of the galaxies that harbor black holes.
After estimating the characteristic mass $M_g$ of the general galaxy
population from its measured correlation length (Adelberger et al. 2005),
we use the GIF-LCDM simulation to calculate as a function of $M_h$ how strongly galaxies
of mass $M>M_g$ cluster around galaxies of mass $M>M_h$.
We infer the masses of the galaxies that harbor various black holes
by finding the value of $M_h$ required to match the observed cross-correlation
length $r_0$.
Figure~\ref{fig:r0_vs_mass} shows the relationship we used to estimate
from our measured cross-correlation length $r_0$
the typical mass of the galaxies containing the black holes (green points).
Adopting other plausible relationships between $r_0$ and $M_h$
would change the inferred masses by less than their random uncertainties.
Percival et al. (2003) and Kauffmann \& Haehnelt (2002)
have shown that halos undergoing mergers have the
same correlation length on large scales as other halos of the same mass,
so our estimates of $r_0$ should provide reasonable
estimates of the halos masses even if AGN are fueled by mergers.

To estimate the random uncertainty in $r_0$, we took a Monte-Carlo approach
that exploited the similarity of the AGN-galaxy cross-correlation length
to the galaxy-galaxy correlation length.  We generated many alternate
realizations of our data by treating randomly chosen galaxies in each field as that
field's AGN, rather than the true AGN themselves,
and recalculated $r_0$ for each simulated sample.  Since the galaxies in our
survey outnumber the AGN by more than twenty to one, the simulated samples
are nearly independent of each other and of the true sample.  We took the
rms spread in $r_0$ among them as the $1\sigma$ uncertainty in our measured 
measured correlation length $r_0^{\rm obs}$.
The distribution of $\chi^2\equiv\sum [(r_0^{\rm obs}-r_0^{\rm pred})/\sigma_{r_0}]^2$
for the predicted values of $r_0$ in figure~\ref{fig:emh_vs_mbh}
should be roughly equal to the distribution
of $\chi^2$ in the simulated samples around the 
line $r_0^{\rm pred} = {\rm constant} = r_0^{\rm gg}$, where $r_0^{\rm gg}$ is the
galaxy-galaxy correlation length in our sample. 
We used this distribution to associate our measured values of $\chi^2$
with a $P$-value.

Our conclusion depends on the assumption that the estimated black hole
masses $M_{\rm BH}$ are not wildly inaccurate.  
We estimate $M_{\rm BH}$ from an AGN's luminosity $l\equiv\lambda L_\lambda$
at $\lambda=1350$\AA\ and CIV line-width ${\rm FWHM}$ with
the relationship that is observed in the local universe:
$M_{\rm BH}/M_\odot\simeq 10^{6.2} (l/10^{44} {\rm erg\, s}^{-1})^{0.7} ({\rm FWHM}/1000 {\rm km\, s}^{-1})^2$ (Vestergaard 2002).
Correcting for a stellar contribution
to the AGNs' luminosities (which we have not done)
would decrease our lowest observed values $M_{\rm BH}$ even further,
strengthening our conclusions.  Our estimated black-hole masses would be
too low for some AGN with small $M_{\rm BH}$ if their observed CIV emission line
were produced in the
narrow-line region rather than the broad-line region (as we assume).
In this case the line widths would be roughly equal to the galaxies'
stellar velocity dispersions (Nelson 2000), at least for radio-quiet AGN,
but in fact the galaxies' mean stellar
velocity width ($\sim 200$ km s$^{-1}$) is an order of magnitude smaller than
the mean AGN line-width for $M_{\rm BH}<10^8 M_\odot$
($2100$ km s$^{-1}$) or $M_{\rm BH}>10^8 M_\odot$ ($4900$ km s$^{-1}$).
It is far smaller than even the smallest observed 
AGN line-width in our sample, $800$ km s$^{-1}$.
Radio-loud AGN make up too small a fraction of our sample to
affect our results if omitted.  
In any case, the observed range
of $M_{\rm BH}$ is so large that our estimates of $M_{\rm BH}$
would have to be wrong by $\sim 1$ order of magnitude to alter our
results significantly.  We cannot rule out the idea that
the relationship between $M_{\rm BH}$, luminosity, and line-width was
utterly different in the past, but it seems easier to believe
that the relationship between $M_{\rm BH}$ and $M_h$ has not changed at all.

\end{document}